# Spectrum Analysis with the Prime Factor Algorithm on Embedded Systems

Josh Vernon and D.G. Perera

Department of Electrical and Computer Engineering, University of Colorado Colorado Springs

Colorado Springs, Colorado, USA

**Abstract**

This paper details the purpose, difficulties, theory, implementation, and results of developing a Fast Fourier Transform (FFT) using the prime factor algorithm on an embedded system. Many applications analyze the frequency content of signals, which is referred to as spectral analysis. Some of these applications include communication systems, radar systems, control systems, seismology, speech, music, sonar, finance, image processing, and neural networks. For many real-time applications, the speed at which the spectral analysis is performed is crucial. In order to perform spectral analysis, a Fourier transform is employed. For embedded systems, where spectral analysis is done digitally, a discrete Fourier transform (DFT) is employed. The main goal for this project is to develop an FFT for a 36-point DFT on the Nuvoton Nu-LB-NUC140V2. In this case, the prime factor algorithm is utilized to compute a fast DFT.

# Section 1: Introduction

This paper details the purpose, difficulties, theory, implementation, and results of developing a fast Fourier transform using the prime factor algorithm on an embedded system. In this introduction, four sections will be examined: (i) an example scenario, (ii), the problem description, (iii) the motivation for the project, and (iv) the main goal of the project.

Scenario: In many applications we wish to analyze the frequency content of signals. This is referred to as spectral analysis. The applications include but are not limited to communication systems, radar systems, control systems, seismology, speech, music, sonar, finance, image processing, neural networks, etc. Clearly, this is a large and broad list. There are almost an endless number of scenarios that frequency analysis can be used in. For example, a radar system can use spectral analysis to estimate the velocity of an object by measuring the object's Doppler shift. A communication system can use spectral analysis to verify signal bandwidth and carrier frequency. Spectral analysis can be used to separate different instruments in a song. The list of applications is far from few.

The Problem Description: In real-time applications, the speed at which the spectral analysis is performed is crucial. If it is not computed fast enough, the system could break. In the previous radar example, the rate at which the velocity is estimated is proportional to the resolution of those estimates. Furthermore, the tracking system could lose a target if the target's position changes quickly, and the estimates are not updated at a quick enough rate. Therefore, it is clear that the problem can be described as *fast* spectral analysis.

In order to perform spectral analysis, a Fourier transform is employed. In an embedded system, the spectral analysis happens digitally. Therefore, a *discrete Fourier transform* (DFT), the Fourier transform's counterpart, is required. A DFT is defined by the following equation:

$$X[k] = \sum_{n=0}^{N-1} x[n] W_N^{nk}, \text{ where } W_N = e^{-j2\pi/N}.$$

The DFT is an orthogonal transform that can be used to find the discrete frequency content of the input signal, $x[n]$. As mentioned above, this can be useful in an abundance of scenarios. When it comes to real-time systems that require the DFT to be computed with low latency and high throughput, the equation shown above is not practical for many scenarios. The output of the DFT is a column vector that is $N$ complex samples in length. To compute one of these output samples, $N$ complex multiplications are required between $x[n]$ and $W_N^{nk}$. (One complex multiplication is equivalent to four real multiplications and two real additions.) Therefore, in order to process the entire DFT, $N \times N = N^2$ complex multiplications are required. The Big O time complexity of this transform is therefore $O(N^2)$.

On an embedded system, applications are running in real-time. This means that DFTs deployed on an embedded system need to be able to perform spectral analysis in real-time as well. This can be a problem if the application requires spectral analysis at a rate that the embedded system cannot keep up with due to the time complexity of the DFT. It is desirable to find mathematically equivalent alternatives to computing the DFT equation directly. The *fast Fourier transform* (FFT) describes a set of algorithms that do exactly this.

      <u>Motivation for the Project</u>: My personal motivation for this project is to explore several FFT algorithms. As a digital signal processing engineer, FFTs are something that I come across on a regular basis. In a non-real-time application such as system modeling, these FFTs can be computed via MATLAB and pose no real problems. However, when the time comes to implement a model onto an embedded system, the need for efficiency becomes very high. Throughput requirements can be very stringent, and the FFT required in the processing is often the bottleneck. Exploring several styles of FFT algorithms is critical in finding an optimal implementation based on the design requirements.

      <u>Main Goal</u>: The main goal for this project is to develop an FFT for a 36-point DFT on the Nuvoton Nu-LB-NUC140V2. The system will utilize the *prime factor algorithm* (PFA) to compute a fast DFT. After the algorithm is developed and coded, the on-board ADC will be initialized in differential input mode to continuously sample an analog input generated by the Keysight waveform generators that are in the lab. Every 1/2 of a second, the software will read 36 samples from the ADC. The digitized waveform will then be passed through the 36-point FFT. The output of the FFT will be written to the local computer in the lab via UART. MATLAB will be used to read in the serial data. The FFT samples will then be converted to a magnitude and interpolated. The maximum value will be used to estimate the input signal's voltage and frequency. These results will be plotted and updated every 1/2 of a second.

## Section 2: Project Description

      This project will communicate the effectiveness of the PFA through the report and will demonstrate its functionality through the hardware setup. The full system overview is shown in Figure 1. The in-lab signal generator will be used to generate a sinusoidal input to the differential input of the NUC's ADC. The digitized waveform will be transformed by the PFA FFT and written to MATLAB via UART for spectral analysis. Because one of the goals of this project is to explore the PFA's effectiveness in reducing the time complexity of a DFT on an embedded system, this section will examine the architecture and computational savings of the PFA-36.

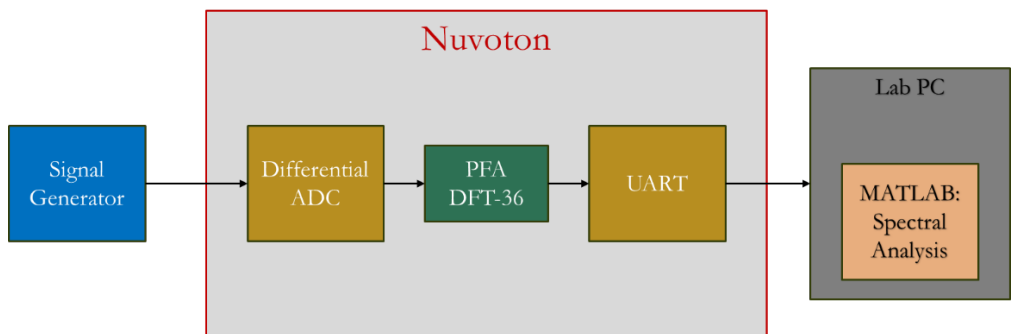

**Figure 1**: System Overview

Most FFTs reduce the time complexity to $O(N\log N)$ by utilizing the fact that $W_N^{nk}$ is symmetric and cyclic. This allows redundant calculations to be omitted. The most commonly used FFT is the Cooley-Tukey algorithm. This is due to its divide-and-conquer approach of breaking down the DFT into smaller subcomponents. This algorithm is however restricted to powers-of-two-sized DFTs, i.e., $N = 2^\alpha$ where $\alpha$ is a positive integer. Not as commonly used is the *Winograd Fourier transform algorithm* (WFTA). The WFTA is more complex to derive and implement, but it is not restricted to powers-of-two. This becomes attractive when a prime-sized FFT is desired. The final FFT that is considered in this project is the *prime factor algorithm* (PFA). The PFA also has a complex derivation and design relative to the Cooley-Tukey algorithm, but it works very well on composite-sized DFTs. A composite number is one whose factors are relatively prime. This algorithm works by breaking up the composite-sized DFT into smaller FFTs that are the size of the factors, and then stitching them together appropriately. The goal is to find an FFT for the 36-point DFT. Because 36 has factors 9 and 4 which are relatively prime, the PFA can be used to compute the FFT-36.

The first step then is to find a 4-point FFT. Based on the previous discussion, the Cooley-Tuley algorithm can be used here because 4 is a power-of-two. Figure 2 shows the block diagram for the 4-point FFT. This architecture is comprised of 8 complex additions and two complex multiplies upon inspection. However, the two complex multiplies are not required. The $W_4^0$ factor equates to 1, so no operation is required here. The $W_4^1$ factor equates to $-j$, and this multiply can be accomplished simply by swapping the input's real and imaginary parts, and then negating the imaginary part. This results in 0 complex multiplies for this block, giving a total of 16 real additions and 0 real multiplies.

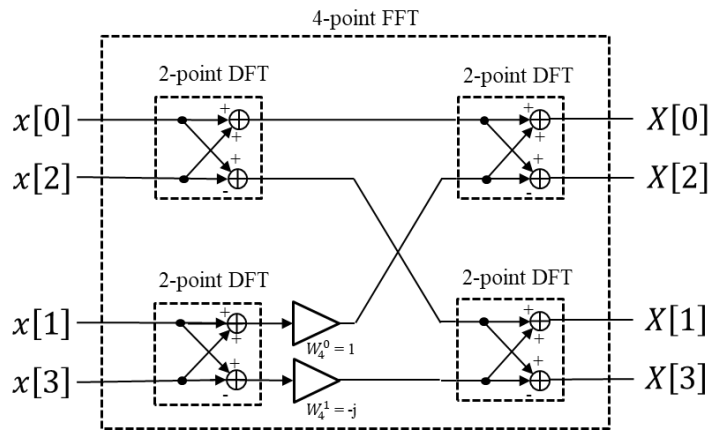

**Figure 2**: 4-Point FFT

The next step is to find a 9-point FFT. As mentioned previously, the Cooley-Tukey algorithm is restricted to powers-of-two, so it will not be a suitable algorithm for this DFT-9. The WTFA *can* accomplish a 9-point FFT; however, this is not the approach that is taken. Instead, the 9-point FFT will be computed by stitching several WTFA-3's together which is slightly more efficient. To compute the WFTA-3, the following equations are used, where $u = \pi/3$. A total of 6 complex additions and 2 complex multiplies are required. This results in a total of 16 real additions and 8 real multiplies.

$$t_1 = x[1] + x[2], \quad t_2 = x[1] - x[2]$$
$$m_0 = x[0] + t_1, \quad m_1 = (\cos(u) - 1)t_1, \quad m_2 = j\sin(u)t_2$$
$$s_1 = m_0 + m_1$$
$$X[0] = m_0, \quad X[1] = s_1 - m_2, \quad X[2] = s_1 + m_2$$

The 9-point FFT can then be created by stitching the WFTAs (also referred to as WFFTs) together as shown in Figure 3. The total computation for this FFT is the combined sum of the WTFA-3's with the additional $W_N$ factors in between stages. This equates to 36 complex additions and 16 complex multiplies, which gives a total of 104 real additions and 64 real multiplies.

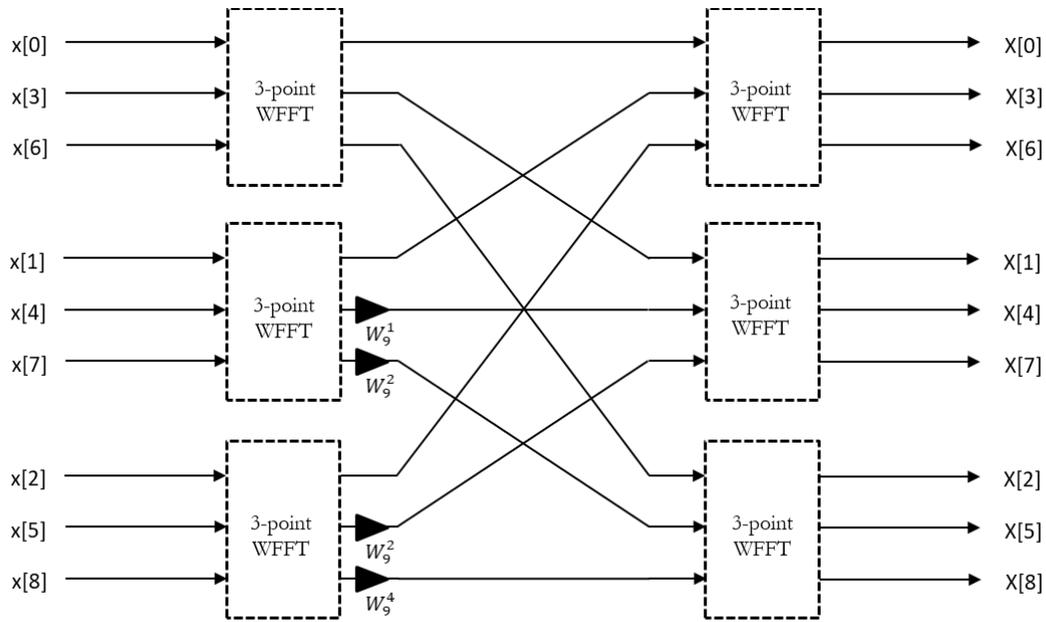

**Figure 3**: 9-Point FFT

Now that the FFT-4 and the FFT-9 structures have been developed, the FFT-36 can be constructed. The PFA requires four 9-point DFTs and nine 4-point DFTs. The complex part of the PFA architecture is the input and output permutations. Figure 4 shows the block diagram for the final PFA FFT. A very appealing result of the PFA is the lack of $W_N$ factors in between the FFT stages. While still considered to be $O(N\log N)$, the PFA is very efficient due to this result. The total computation is the combined sum of the FFT-9s and the FFT-4s. Therefore, the total number of complex additions is 216, and the total number of complex multiplies is 64. This results in a total of 560 real additions and 256 real multiplies.

These final results can be compared with the original DFT equation. As previously stated, the DFT equation requires $N^2$ complex multiplies. For the DFT-36, this equates to 1296 complex multiplies, which results in 2592 real additions and 5184 real multiplies. Therefore, the PFA results in a 78.4% reduction of real additions and a 95.1% reduction in real multiplies!

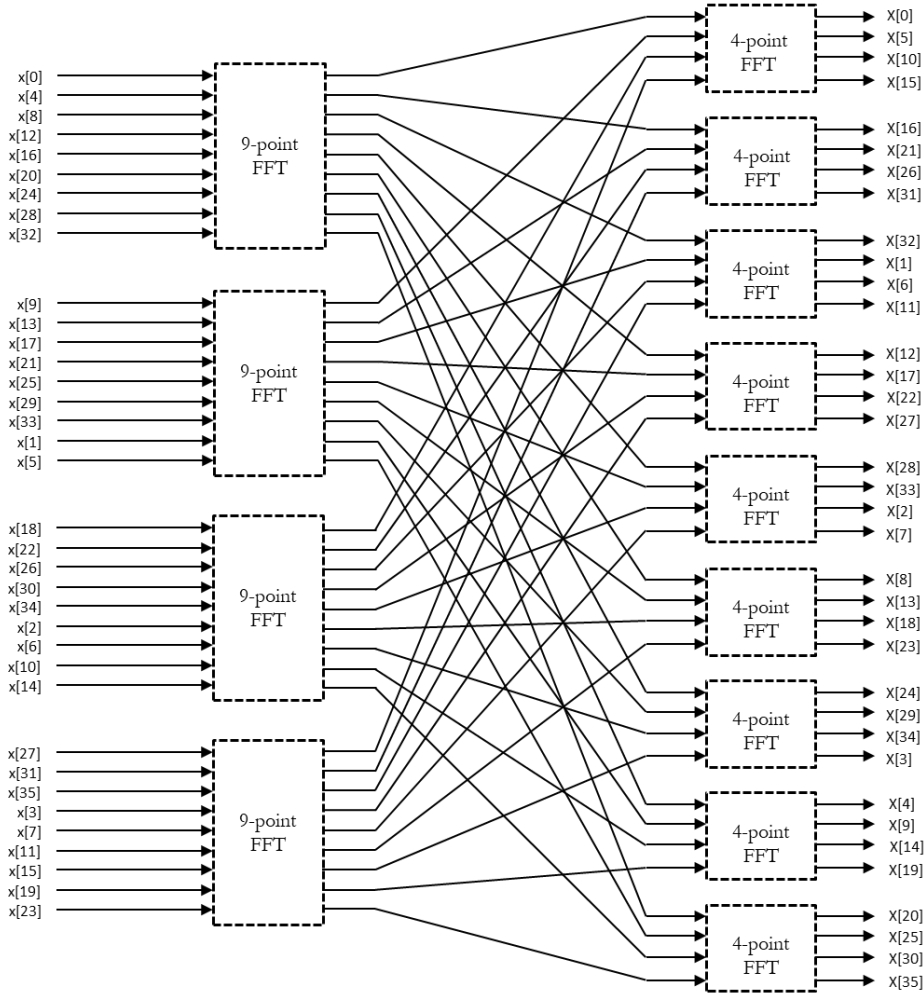

**Figure 4**: 36-Point FFT

## Section 3: Hardware Design

The hardware design for the project is rather straightforward and can be seen in Figure 5. The Keysight InfiniiVision MSOX4024A is used as both a signal generator and an oscilloscope. The output of the signal generator is connected to both the oscilloscope and the NUC ADC. The purpose of the oscilloscope is to ensure that the specified parameters to the signal generator are accurate. The feed between the signal generator and two sinks is a coaxial cable. A coax-to-clip connection is used to connect to the NUC GPIO pins.

The differential mode of the ADC requires two inputs: $V^+$ and $V^-$. The ADC channels are paired to the GPA GPIO pins, and when in differential input mode, the 8 ADC channels are used in tandem to create 4 differential ADC channels. The even indexed channel is $V^+$ and the odd-indexed channel is $V^-$. For this project, ADC channel 6 is being used and therefore so is channel 7.

The other hardware component is the UART connection between the NUC and the lab PC. This is a straightforward connection by UART-to-USB. Although spectral analysis occurs through the MATLAB *software*, it is an external component to the system, so it will be discussed briefly here. The MATLAB software was chosen for this design because it can accomplish many tasks in a single script. Firstly, it is used to read the COM port as floating point data. That data is then converted into a complex number datatype, and the magnitude of the 36 samples is taken. Those magnitudes are then interpolated across 512 discrete frequencies. The largest interpolated magnitude is measured, and the voltage and corresponding frequency are stored. The double-sided spectrum is then plotted in the form of the FFT bins and the interpolation thereof. The script then waits for the next FFT frame to arrive so that it can recompute and update the plot.

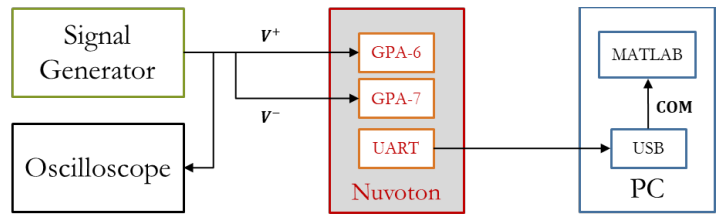

**Figure 5**: Hardware Diagram

This hardware setup was chosen because I felt that it provided a reasonable way to demonstrate the functionality and versatility of the FFT algorithm. No problems were encountered in the hardware development for this project.

## Section 4: Software Design

The software design for this project makes an attempt to be effective and efficient through the C language and file structure. All functions have their own header files which include the function definition as well as any dependencies. A structure datatype is created to store complex numbers as well. The FFT design is implemented in floating point, so the structure contains two floating point elements: one for the real part and the other for the imaginary part. This structure is defined in its own header file because it is a general datatype that can be used throughout multiple functions in the code. The ADC channel, maximum bytes per UART send, digital-to-voltage conversion value, and FFT size $N$ are all defined as preprocessor values.

A complex multiply function is created to handle all required complex multiplication operations. This function has three inputs that are passed by reference: the first complex float in the multiplication $a$, the second complex float in the multiplication $b$, and the resulting complex float product $c$. This function performs the multiplication by the following equation:

$$c_{real} = a_{real}b_{real} - a_{imag}b_{imag}$$
$$c_{imag} = a_{real}b_{imag} + a_{imag}b_{real}$$

As mentioned, all components are passed by reference in this operation. This is not required, but it helps simplify the code that calls this function.

Functions are created for the 2-point FFT, the 4-point FFT, the 3-point WFTA, the 9-point FFT, and the 36-point PFA. Each of these functions has one input and one output. The input is a pointer to the memory address of the first element in the $N$-element array that is to be processed by the $N$-point FFTs. These memory addresses are referenced as the real part of the complex float structure datatype. Therefore, each pointer increment references 64 bits – 32 bits from the real part and 32 bits from the imaginary part. These FFTs are essentially coded to the architecture shown in Figures 2, 3, and 4. The FFT-2 only performs complex addition/subtraction and does not call the complex multiply function. Using $u = \frac{\pi}{3}$, the WFTA-3 pre-defines $\cos(u) - 1$ and $\sin(u)$ since these values do not change. They are defined as $\cos(u) - 1 = -1.5$ and $\sin(u) = 0.8660254$. This function performs 6 real additions and two real subtractions. The complex multiply function is called twice in this function. The FFT-4 function calls the FFT-2 function four times. Instead of multiplying by complex number by $-j$, the function swaps the real and imaginary components of the complex number and then negates the imaginary component. Therefore, this function does not need to call the complex multiply function. The FFT-9 defines the following factors that are seen in Figure 3: $W_9^1 = 0.7660444 - j0.6427876$, $W_9^2 = 0.1736482 - j0.9848775$ and $W_9^4 = -0.9396926 - j0.3420201$. This function calls the WTFA-3 function six times and calls the complex multiply function four times. The PFA-36 function has no $W_N$ factors in between stages. It calls the FFT-9 function four times and the FFT-4 function nine times. This function primarily stitches the FFT-4s and the FFT-9s together and performs the appropriate input and output index permutations.

The main function of the program defines a union datatype. The union allows memory addresses to be referenced as different datatypes. This is beneficial in this system because the UART Write command expects a uint8_t pointer when referencing data to write. Previously, the workaround for this was to call the sprintf function to convert numeric values to ASCII characters and then cast the char array to uint8_t. This worked in previous systems because only a few integer-type values were being sent over UART at a time. In this system, floating-point is being used. This means that up to seven fractional numerals can be present at a time. Furthermore, in this design, there are up to three integer numerals as well. This equates to a total of ten numerals and therefore ten ASCII characters would be required to send per floating point value. With 36 complex floating point values per frame, which is a total of 72 floating point values, this would require 720 bytes of information to be delivered over UART. Furthermore, a lot of latency would accrue from calling sprintf on 72 floating point values. When a union is created to access memory as both the complex float structure *as well as* uint8_t, the raw data from memory can be transferred via UART. This eliminates the latency overhead from the sprintf call. Furthermore, only four bytes are required to send a floating point value with this method. This means that only 288 bytes are required to send over UART per frame, resulting in a 60% increase in data transfer efficiency.

The main function starts by defining two unions as defined above. The first union is for the input data. This array does not necessarily need to be referenced as a uint8_t but declaring it as the union datatype doesn't impede the processing at all. Another note is that this input array also isn't complex in nature since the sampled data is purely real but defining it as complex allows the FFT functions to be coded in a more programmatic way. The second union is for the output data, which is complex and does need to be referenced as a unint8_t. Both unions are initialized to zero. A char array is created for LCD printing. The initialization structure for the UART, a looping variable, a sample rate variable, and two variables that determine UART Write sizes are also defined.

As done in all previous labs, the system clock is set up with the 22 MHz oscillator. Next, the LCD is initialized and cleared using the LCD driver functions. A timer is then set up to compute DFT-36 frames on an interval with the use of an ISR. Timer 1 with clock source 7 is selected in periodic mode with 10 ticks per second. The ADC is then set up. ADC channel 6 is selected, and therefore channel 7 is used as the negative input. These are pins GPA-6 and GPA-7, respectively. Using the GPIO driver function, both digital inputs on these pins are disabled. Using the ADC driver function, the ADC is opened in differential and continuous mode using the 22 MHz internal clock. Also using an ADC driver function, the output format of the ADC is set to two's complement in order to receive signed data. Next, the UART is set up with clock source 7. UART0 is used because it can transfer up to 64 bytes at a time. The baud rate is set to 9600, the number of data bits is set to 8, the number of stop bits is set to 1, no parity bits are used, and the FIFO triggers at 1 byte.

At this point, all of the initialization and set up has been accomplished. The next step is to return the ADC conversion rate and print it to the LCD using the ADC and LCD driver functions. Given that there are 36 complex samples in a frame, and that UART can transfer up to 64 bytes per write, the total number of writes per frame and residual bytes to send is computed. The total number of 64-byte writes per frame is evaluated as $\lfloor (4 \times 2N)/64 \rfloor = 4.5$. The residual number of bytes to send evaluates to $64[(4 \times 2N) \bmod 64] = 32$. From here, the ADC is self-calibrated. Once it is done doing so, it is instructed to begin ADC conversions.

This is the part of the program that begins its infinite loop. The first thing that happens is a check to see if the timer ISR has updated. When the ISR counts 5 times (0.5 seconds), a flag is asserted. When that flag is not asserted, nothing happens, and the program will continue to check. When the flag is asserted, then the flag is reset, and the data processing occurs. A for-loop iterates 36 times to read ADC values, convert them to floating point, and then convert them to a voltage value. (According to the TRM, the data is converted in differential input mode by $V_{diff} = V^+ - V^-$.) These converted values are stored in the input array. The $0^{th}$ element address of the array is passed into the PFA-36 function along with the $0^{th}$ element of the output array. The output array is then set in 5 iterations – four of which are sending 64 bytes at a time and the last of which is sending the remaining 32 bytes. This completes the main program.

To test the code, primarily the PFA, each segment was broken down and tested individually. To start, test vectors were defined in the main program to check the FFT-2, FFT-4, WFTA-3, FFT-9, and PFA-36. These test vectors were generated in MATLAB, where the true DFT output was computed for the C verification. The Keil debugging tool was utilized first to catch errors, and then the UART Write was utilized to ensure that the data was being sent to the COM port and being read properly. There were a few instances where the FFTs were coded incorrectly (mostly due to the nontrivial permutations of the PFA), so the debugging proved to be useful. To wrap this section up, Figure 6 is provided to outline the software processing.

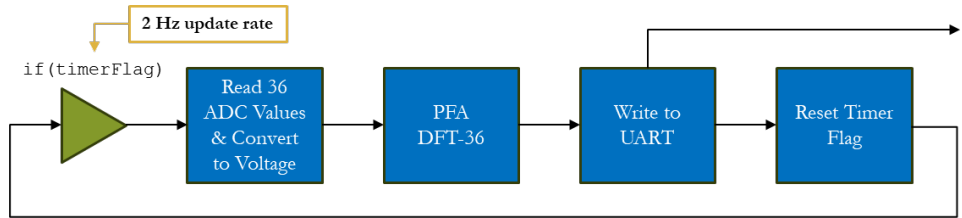

**Figure 6**: Software Flow Diagram

## Section 5: System Design

The integration between hardware and software was not very complicated and can be seen in Figure 7. On the output side, the biggest hurdle was transmitting data over UART efficiently. This was achieved (as mentioned previously) by using the union datatype to access floating point data as uint8_t data. On the input side, there were a couple of confusions. Firstly, the differential ADC appears to produce some oddities.

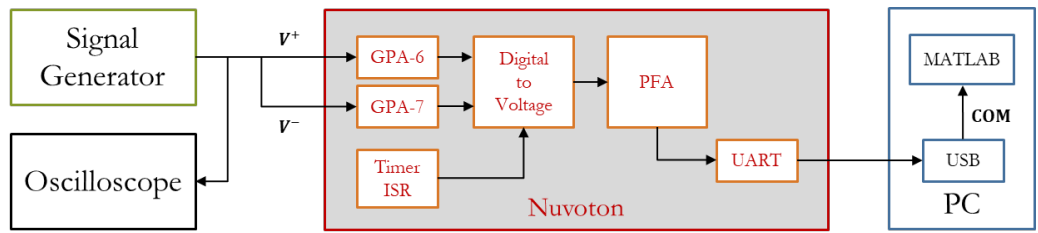

**Figure 7**: System Diagram

Firstly, the sample rate that is returned from the ADC driver function did not match the observed sample rate. The TRM states that the ADC takes 27 clock cycles to convert one sample, and the driver function accounts for this. However, manually tuning the signal generators tonal frequency revealed that folding would occur around 21 MHz. Because of this observation, the MATLAB plots were adjusted to use this frequency range. Another confusion is with respect to a DC spike that would appear when the input was simply a sinusoid. The DC spike vanishes when an offset around 100 mV is applied to the tone. Another notable issue was the voltage conversion. The value was incorrect, but this was because the two's complement mode loses 1 bit of resolution due to the sign bit, and the maximum voltage range is doubled from 3.3V to 6.6V. After making these updates, the voltage value was corrected. Finally, the biggest issue was the ISRs. Both a heartbeat and a timer ISR were implemented, but upon running the program, they would both freeze. When commenting out the FFT portion of the code, the ISRs began working again. Upon checking the debugger, it seems as though the FFT is utilizing too many resources for the CPU to handle. To overcome this, no heartbeat function is used and the timer ISR is converted into a timer delay function.

## Section 6: Discussion, Conclusion, and Future Work

Overall, the project was a success. The prime factor algorithm utilized the fact the DFT was composite in size, and utilized the Cooley Tukey algorithm and the Winograd Fourier transform for its subcomponents. Overall, the PFA was able to reduce the number of DFT real additions by 78.4% and the number of real multiplications by 95.1%. The final product was able to sample an analog waveform using the differential ADC, compute a fast and efficient discrete Fourier transform, write the output to UART efficiently, and perform real-time spectral analysis. The voltage and frequency estimation results are reasonably accurate as well. Figure 8 shows a single frame of the spectral analysis produced from the FFT on the NUC in tandem with the MATLAB code.

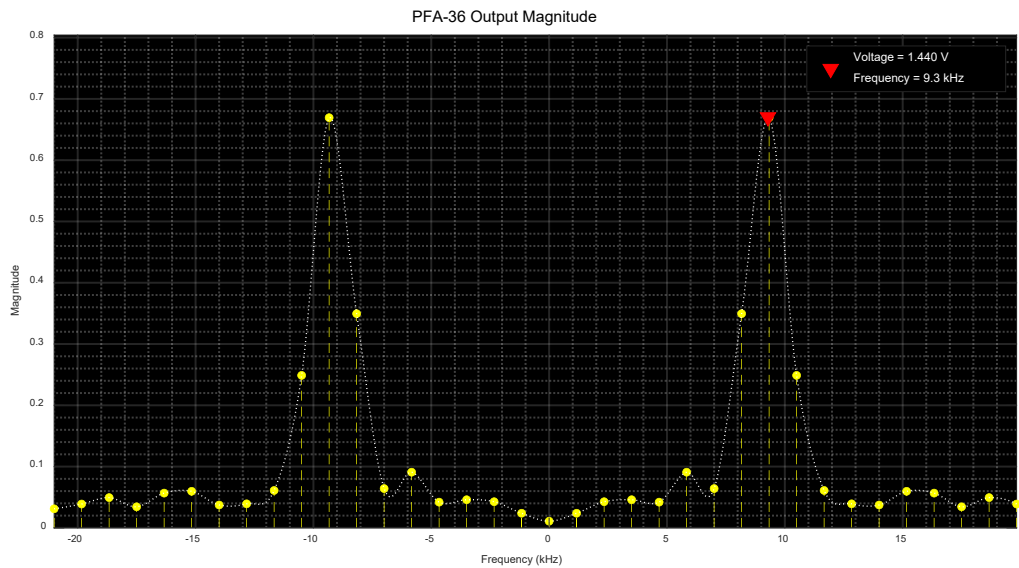

**Figure 8**: MATLAB Output

While this project was a success, there are several areas that could be improved in the future. Firstly, while the PFA does reduce the number of computations performed, the current implementation probably utilizes more memory than it needs to. This could and should be reduced for a project that requires careful memory allocation. Another interesting area to explore would be to push the timer ISR to the limit, i.e., determine the fastest update time given the latency of the PFA and UART transmission. This could be used for high-speed applications. If the UART is taken away, the voltage and frequency estimation could happen directly on the NUC and be printed to the LCD. On the hardware side of things, it would be worthwhile to dive deeper into the workings of the differential ADC and understand why the 100mV offset was needed to remove the DC component. Another important task would be understanding why the sampling rate did not align with the observed sampling rate is critical to any real application that requires frequency estimation on the NUC itself. Whether it was due to clock cycle delays from the for-loop or just an oversight, it should be tracked down. Last but not least, determining how to get the ISRs working on the board while the FFT is running.

This work is inspired by the embedded systems and digital design research group at UCCS. This group has done extensive work in embedded hardware and software architectures, techniques, and associated models. Their analyses [7],[8] demonstrate that FPGA-based embedded systems are currently the best option to support complex applications and algorithms, such as the ones presented in this report, i.e., Fast Fourier transform using the prime factor algorithm. Also, their previous work on FPGA-based embedded accelerators, architectures, and techniques for various compute and data-intensive applications, including data analytics/mining [9],[10],[11],[12],[13],[14],[15],[16],[17],[18]; control systems [19],[20],[21],[22],[23],[24]; cybersecurity [25],[26],[27]; machine learning [28],[29],[30],[31],[32],[33]; communications [34]; edge computing [35],[36]; bioinformatics [37]; and neuromorphic computing [38],[39]; demonstrated that FPGA-based embedded systems are the best avenue to support and accelerate complex algorithms. Therefore, as future work, we are planning to create FPGA-based hardware for Fast Fourier transform using the prime factor algorithm.

For the aforementioned FPGA-based hardware architectures for Fast Fourier transform using the prime factor algorithm, we are planning to investigate hardware optimization techniques, such as parallel processing architectures (similar to [30],[40],[41],[42]), partial and dynamic reconfiguration traits (as stated in [43],[44],[45]) and architectures (similar to[26],[46],[47],[48]), HDL code optimization techniques (as stated in [49],[50]), and multi-ported memory architectures (similar to [51],[52],[53],[54]), to further enhance the performance metrics of FPGA-based Fast Fourier transform using the prime factor algorithm, while considering the associated tradeoffs.